# Multi-ring necklace vortex solitons in Kerr nonlinear media with azimuthally modulated Bessel potentials

**RUOLAN ZHAO,[1] JING CHEN,[1] BORIS A. MALOMED,[3,4] AND RONGCAO YANG [1,2,*]**

[1] *College of Physics and Electronics Engineering, Shanxi University, Taiyuan 030006, China*
[2] *Shanxi Key Laboratory of Wireless Communication and Detection, Taiyuan, 030006, China*
[3] *Department of Physical Electronics, School of Electrical Engineering, Faculty of Engineering, Tel Aviv University, Tel Aviv 69978, Israel*
[4] *Instituto de Alta Investigación, Universidad de Tarapacá, Casilla 7D, Arica, Chile*
* *sxdxyrc@sxu.edu.cn*

**Abstract:** We address the existence, stability, and dynamics of single-ring and multi-ring vorticity-carrying necklace solitons under the action of the Kerr nonlinearity and a Bessel-lattice potential modulated in the azimuthal direction. The model may be realized in the spatial domain for bulk optical waveguides, the spatiotemporal domain for optical cavities, and for effectively two-dimensional Bose-Einstein condensates. The setup supports single- and multi-ring necklace vortex patterns, including monopoles, dipoles, tripoles, quadrupoles, pentapoles, sextupoles, octupoles, and 12-poles. In contrast with the inherent instability of conventional vortex beams with high topological charges (winding numbers), vortex necklace-shaped solitons with large winding numbers are found to be stable in the present setup. In particular, octupoles exhibit stable breathing dynamics, and 12-pole necklaces with high winding numbers may be stable. These findings provide a new way for generating stable vortex necklaces, offering a vast potential for manipulations of complex spatiotemporal light fields.

## 1. Introduction

The creation and use of optical vortex beams (OVBs) have drawn a great deal of interest in nonlinear photonics [1, 2], ultracold Rydberg atomic gases [3, 4], and Bose-Einstein condensates (BECs) [5, 6]. OVBs are characterized by a helical wavefront, central phase singularity, and the ability to carry quantized orbital angular momentum (OAM) [7]. They provide the transfer of OAM to microscopic particles, markedly promoting the development of the optical-tweezers technology [8, 9], and find broad applications in image encryption [10], super-resolution recognition [11], optical communications [12, 13], and quantum-information processing [14-16]. While traditional OVBs were predicted and created in the spatial domain, as stationary beams carrying OAM in bulk optical media [17], the tilt of the vortex axis in the space-time plane gives rise to *spatiotemporal vortices* [18-25].

However, the propagation of vortex beams in nonlinear media faces significant challenges, as concerns their stability [26, 27]. In effectively two-dimensional (2D) self-focusing cubic (Kerr) nonlinear media, the beam collapse ensues once the input power exceeds the critical threshold, while below this threshold, it suffers decay under the action of diffraction [28-30]. Still stronger is the instability of the vortices against azimuthal perturbations, which split them into a set of separating fragments [26, 27, 31]. The splitting instability impedes the application of OVBs in nonlinear optics.

Two basic methods have been broadly adopted to overcome the OVB instability. One approach relies on the use of competing self-focusing and defocusing nonlinear effects [32-35]. In particular, the cubic-quintic nonlinearity supports stable vortex solitons with unitary and multiple values of the winding number [32, 36-38], and the quadratic-cubic combination of nonlinear terms provides for the stability of quadrupole and octupole solitons [39, 40]. Another remarkable example is the predicted stabilization of vortex quantum droplets under the action of the competing mean-field self-attractive cubic nonlinear term and the repulsive Lee-Huang-Yang (LHY) quartic correction in BECs [41-43]. Another alternative approach for suppressing the azimuthal instability of vortex solitons is to introduce an external axisymmetric potential [44-46]. In particular, stable higher-order vortex solitons have been realized in cubic-quintic nonlinear media with the harmonic-oscillator potential [47] and Kerr nonlinear media with the flat-bottom potential [48]. The interplay of radially periodic potentials with the cubic-quartic or cubic-quintic nonlinearity gives rise to stable 2D nested vortex solitons with different topological charges [49, 50]. Various stable OVB solitons can also exist in other external potentials, such as ring vortex solitons maintained by ring-shaped potentials [51], necklace vortex solitons in discrete waveguiding arrays [52], vortex light bullets in quasi-phase-matched photonic crystals [53], multipole vortex solitons in moiré superlattices [54], etc.

In particular, Bessel lattices and Bessel beams, due to their axial symmetry and multi-ring non-diffractive structural characteristics, provide significant advantages in the creation of OVB solitons [55-57]. Soliton clusters exhibiting orbital motion in different lattice rings were investigated in cubic nonlinear media [58]. Tilted discrete solitons were experimentally generated by modulated fragmented Bessel beams in photorefractive media [59]. Stable 2D and 3D ring-shaped vortex solitons were also discovered in Bessel lattices [60, 61]. However, the existence and stability of complex structured light (in particular, OVB solitons), in azimuthally modulated Bessel lattices, that incorporate local potential barriers and wells, remains unexplored.

This paper aims to investigate the existence, stability and dynamics of diverse necklace OVB solitons in Kerr nonlinear media with azimuthally modulated Bessel potentials. Single-ring and multi-ring necklace vortex solitons can be generated in local Bessel-potential wells, including monopole, dipole, tripole, quadrupole, pentapole, sextupole, octupole, and 12-pole solitons. It is found that the topological charge exerts an abnormal *stabilizing effect* on vortex solitons in Bessel lattices with azimuthally modulated parameters. These results provide new possibilities for the realization of the stable propagation and precise control of complex structured light fields in nonlinear optical systems.

## 2. The theoretical model

The propagation of optical beams along optical axis $z$ in the Kerr nonlinear medium is governed by the nonlinear Schrödinger equation (NLSE). In the cylindrical coordinate system $(r, \theta, z)$, it is written as [46]

$$i\frac{\partial \Psi}{\partial z} = -\frac{1}{2}\left(\frac{\partial^2}{\partial r^2} + \frac{1}{r}\frac{\partial}{\partial r} + \frac{1}{r^2}\frac{\partial^2}{\partial \theta^2}\right)\Psi - |\Psi|^2 \Psi + pV\Psi, \quad (1)$$

where $\Psi$ is the normalized field, $p$ denotes the depth of the adopted potential,

$$V(r,\theta) = V_0(r,\theta) - (V_0)_{\min}, \quad V_0 = J_n\left[(2b_{\mathrm{lin}})^{1/2} r\right]\cos(k\theta), \quad (2)$$

$J_n$ is the $n$-order Bessel function, $b_{\mathrm{lin}}$ is the transverse scale factor determining the lateral extent of the potential, and $k$ is the azimuthal modulation parameter. Here the Bessel-lattice potential parameters are set as $n = 1$, $b_{\mathrm{lin}} = 1$, and $p = 30$. In optics, this effective potential can be burnt into the bulk waveguide by the laser tool based on passing the processing laser beam through a spatial light modulator, combined with an appropriate optical wave plate [62].

While Eq. (1) is written in the spatial domain, its *spatiotemporal counterpart*, with the evolution variable $z$ replaced by time $t$, is relevant too, as a model of the temporal light evolution in a cavity. Furthermore, the same spatiotemporal model represents the Gross-Pitaevskii equation for the mean-field wave function of an effectively 2D BEC, with the external optical-lattice potential induced by a properly shaped broad laser beam illuminating the condensate.

Figure 1 illustrates the shape of the first-order Bessel lattices with different azimuthal-modulation parameters. For $k = 0$ in Eq. (2) (i.e., in the absence of the azimuthal modulation), Fig. 1(a) exhibits a typical isotropic Bessel lattice. Two blue concentric rings in Fig. 1(a) designate circular potential wells where the beam may be trapped. However, in the case of $k \neq 0$, the lattice exhibits a set of concentric fragmented rings, as shown in Figs. 1(b) and 1(c), where the cores circled by the black dotted line are defined as the first ring, which is coaxial with the external ones. Going from the inside to outside, the rings are denoted as $O_n = 1, 2, 3, \ldots$. This concentric lattice features the distribution of alternating potential wells and barriers, as seen in Fig. 1(b). For the deeper azimuthal modulation, the lattice exhibits more complex potential patterns, such as the one in Fig. 1. Such potential landscapes may diversify structured light fields, and suggest new settings for exploring interactions between soliton "beads" that can build necklace structures.

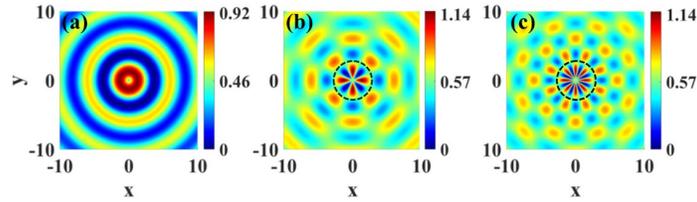

Fig. 1. The azimuthally modulated Bessel lattice defined as per Eq. (2) with $k = 0$ (a), $k = 4$ (b), and $k = 8$ (c).

To seek for stationary OVB solution of Eq. (1), we set

$$\Psi(r,\theta,z) = \phi(r,\theta)\exp(i\mu z), \qquad (3)$$

where $\phi(r,\theta)$ is the stationary OVB field, and $\mu$ is the propagation constant. The topological charge $m$ (winding number) can be defined even for an anisotropic stationary state, as a total phase circulation of the complex field $\phi(r,\theta)$ along a path surrounding the axle, divided by $2\pi$ [44]. Another characteristic of the OVB is its total power, $P = \iint |\Psi|^2 d^2\mathbf{r}$.

Substituting ansatz (3) in Eq. (1) gives rise to the stationary equation,

$$\mu\phi = \frac{1}{2}\left(\frac{\partial^2}{\partial r^2} + \frac{1}{r}\frac{\partial}{\partial r} + \frac{1}{r^2}\frac{\partial^2}{\partial \theta^2}\right)\phi + |\phi|^2 \phi - pV(r,\theta)\phi, \qquad (4)$$

which can be numerically solved by means of the Newton-conjugate gradient method [63].

The stability of the stationary solutions was analyzed by adding a small perturbation to expression (3):

$$\Psi(r,\theta,z) = \left[\phi(r,\theta) + u(r,\theta)\exp(\lambda z) + v^*(r,\theta)\exp(\lambda^* z)\right]\exp(i\mu z), \qquad (5)$$

where $\{u(r, \theta), v(r, \theta)\}$ is the perturbation eigenmode, $\lambda$ is the instability growth rate, and * stands for the complex conjugate. Substituting the perturbed solution (5) in Eq. (1) and performing the linearization with respect to the perturbation, we arrive at the following linear eigenvalue problem:

$$i\begin{bmatrix} M_1 & M_2 \\ -M_2^* & -M_1^* \end{bmatrix}\begin{bmatrix} u \\ v \end{bmatrix} = \lambda \begin{bmatrix} u \\ v \end{bmatrix}, \qquad (6)$$

where $M_1 = \frac{1}{2}\nabla^2 - \mu + 2|\phi|^2 - pV(r,\theta)$, $M_2 = \phi^2$, and $\nabla^2 = \frac{\partial^2}{\partial r^2} + \frac{1}{r}\frac{\partial}{\partial r} + \frac{1}{r^2}\frac{\partial^2}{\partial \theta^2}$ is the transverse diffraction operator. The eigenvalue problem was numerically solved using the Fourier collocation method. OVB solitons are stable only when all eigenvalues in the spectrum are purely imaginary (i.e., Re($\lambda$) = 0). Additionally, results of the stability analysis were verified through simulations of the perturbed propagation, employing the split-step Fourier method.

## 3. Results

To proceed with the analysis, we start from the OVB solitons trapped in the first ring of the Bessel potential with azimuth modulation indices $k$ =1, 2, 3, 4, 5 in Eq. (2). Note that, for $k$ = 1, an off-center monopole soliton is generated, with respect to the location of the potential well, due to the asymmetry of the expression given by Eq. (2) in this case [see Figs. 2(a) and 2(b)]. As $k$ increases, creating more potential wells along the azimuthal direction of the lattice, OVBs with zero vorticity, $m$ = 0, can form dipole [Fig. 2(c)], tripole [Fig. 2(d)], quadrupole [Fig. 2(e)], and pentapole [Fig. 2(f)] solitons, for $k$ = 2, 3, 4 and 5, respectively. Figures 2(g) and 2(h) show the dependence of the power and instability growth rate on the propagation constant, for these soliton families. In Fig. 2(g) and in similar figures presented below [Figs. 3(b), 4(c), 5(a), and 7(a)] the $P(\mu)$ branches commence from $P$ = 0, bifurcating from linear modes, which are produced by Eq. (1) without the cubic term. Note also that the $P(\mu)$ branches in Fig. 2(g) and in their counterparts satisfy the well-known Vakhitov-Kolokolov criterion, which is the necessary (although not sufficient) condition for the stability of the respective soliton families [30, 64].

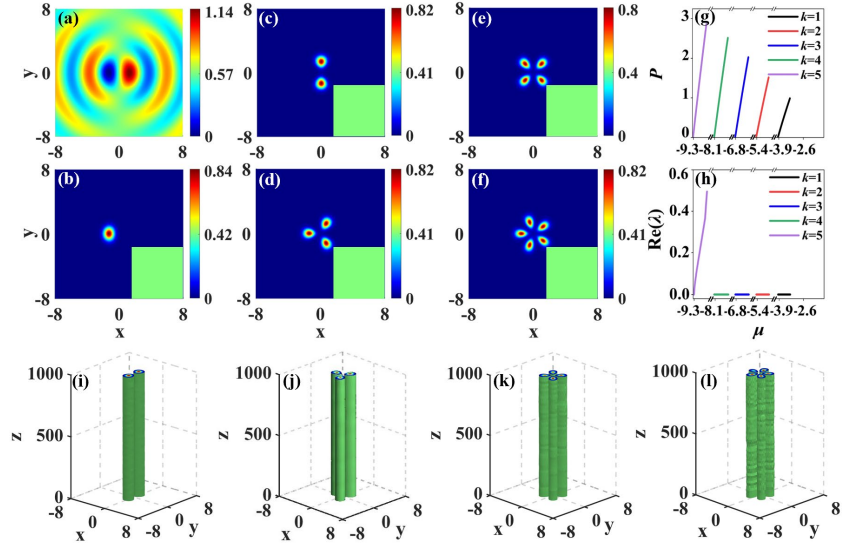

Fig. 2. The field configuration defined as per Eq. (2) with $k$ = 1 (a). The power-density distributions, $|\phi(r,\theta)|^2$, and phases of a stable monopole soliton with $k$ = 1, $\mu$ = -3.6 (b), a stable dipole soliton with $k$ = 2, $\mu$ = -5.1 (c), a stable tripole soliton with $k$ = 3, $\mu$ = -6.5 (d), a stable quadrupole soliton with $k$ = 4, $\mu$ = -7.8 (e), and an unstable pentapole soliton with $k$ = 5, $\mu$ = -9 (f). Panels (g) and (h): the soliton's power $P$ and instability growth rate Re($\lambda$) vs. the propagation constant $\mu$ for the OVB solitons trapped in the first ring of potential (2) with the azimuthal-modulation indices $k$ = 1-5. Panels (i)-(l): the simulated propagation of the solitons displayed in (c)-(f). All the results in this figure pertain to the states with zero winding number, $m$ = 0.

It is seen in Fig. 2(g) that the solitons with more complex structures (higher values of $k$) have larger values of the power for the same propagation constant, because multipole solitons require stronger nonlinearity to localize and stabilize their poles ("beads" of the respective necklace structure). When the soliton power exceeds the critical value the collapse occurs, causing the termination of the $P(\mu)$ curves. Note that the monopole, dipole, tripole, and quadrupole solitons are stable in their entire existence region, in contrast to the unstable pentapoles, see Fig. 2(h). Meanwhile, simulations of the propagation of the solitons for $k = 2$, 3, 4, 5 in Figs. 2(i)-2(l) verifies their stability. It is clear that the dipole, tripole, and quadrupole solitons propagate stably, while the pentapole soliton is unstable. This difference in their stability may be attributed to the relatively sparse arrangement of the "beads" in the monopoles, dipoles, tripoles, and quadrupoles, minimizing the potentially destabilizing effect of the interaction between them.

Figures 3(a)-3(c) depict the power-density distributions, $|\phi(r,\theta)|^2$, of the zero-vorticity sextupole solitons trapped in the first ($O_n = 1$), second ($O_n = 2$), and third ($O_n = 3$) rings with the azimuthal-modulation index $k = 6$ in Eq. (2) and zero winding number, $m = 0$. The dipoles, tripoles, quadrupoles and pentapoles trapped in the rings other than the one closest to the center are not shown here, as the very large separation between their constituent "beads" makes them tantamount to sets of non-interacting constituents. The dependence of the power and instability growth rate of the sextupole solitons, trapped in the three rings, on the propagation constant are presented in Figs. 3(d)-3(f). The total power decreases for the soliton structure confined by the relatively shallow potential trough of the third ring. As the azimuthal-modulation index $k$ in potential (2) increases, i.e., the azimuthal-modulation period becomes smaller, the soliton arrangement in space gets tighter. This leads to stronger interaction between adjacent "beads" and destabilization of the sextupole necklace. Therefore, it is unstable in its nearly entire existence region, see Fig. 3(f). Figures 3(g)-3(i) show the unstable [Fig. 3(g)] and stable [Figs. 3(h) and 3(i)] propagation of the solitons displayed in Figs. 3(a)-3(c).

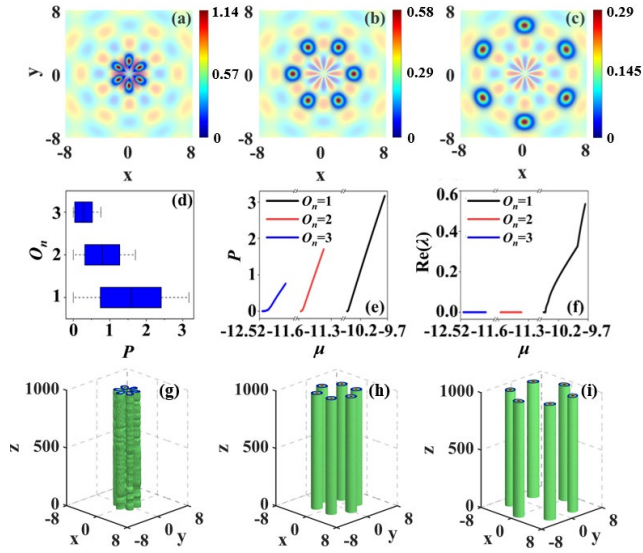

Fig. 3. The superimposed plots of the lattice potential (2) with $k = 6$ and the power-density distribution of the zero-vorticity sextupole solitons with $\mu = -9.8$ (a), $\mu = -11.4$ (b), and $\mu = -12.44$ (c). Panel (d): the existence regions of the sextupoles trapped in different rings. (e) and (f): The power and instability growth rate vs. the propagation constant in the first, second, and third rings. Panel (g)-(i): the unstable and stable propagation dynamics of the sextupoles trapped in different rings.

Proceeding to the presentation of vortex solitons, with nonzero winding numbers, the linear eigenvalue spectrum and linear localized vortex modes trapped in the first ring of potential (2) with the azimuthal-modulation index $k = 6$ are shown in Fig 4(a), where points A and B correspond to the degenerate states for $m = 1$, and points C and D correspond to the degenerate states for $m = 2$, respectively. Thus, the linear localized vortex modes $\phi$ with the topological charges $m = 1$ and $m = 2$ can be obtained as superpositions of the corresponding degenerate states $\phi_A + i\phi_B$ and $\phi_C + i\phi_D$. The degenerate vortex modes are depicted in Fig.4(a). The power-density distributions $|\phi(r,\theta)|^2$ of the vortex solitons trapped in the first ring of potential (2) with $k = 6$ are shown in Figs. 4(b-e) for $\mu = -10.34$ and $-10$. Here the topological charges $m = 1$ and $m = 2$ are considered because their values are limited to $0 < m < k/2$ [65]. The corresponding dependences of the power and instability growth rate on the propagation constant are plotted in Figs. 4(f) and 4(g). It is evident that, at a fixed propagation constant $\mu$, the power increases with the topological charge.

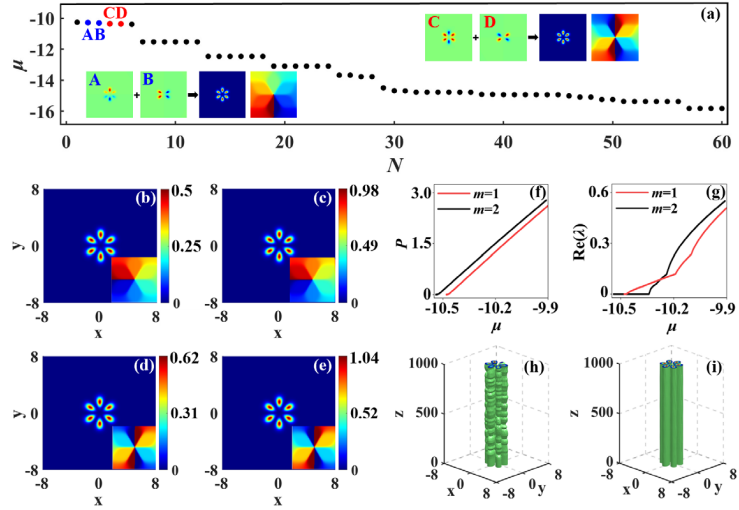

Fig. 4. Panel (a): the linear eigenvalue spectrum and linear localized modes witht $k = 6$. For the components corresponding to $m = 1$ (A, B) and $m = 2$ (C, D), their eigenvalues are represented by blue dots and red dots, respectively, $N$ denoting the eigenvalue number. The power-density distributions and phases of the vortex solitons with propagation constants $\mu = -10.34$ (b) and $\mu = -10$ (c), for $m = 1$. The power-density distributions and phases of the vortex solitons with propagation constants $\mu = -10.34$ (d) and $\mu = -10$ (e), for $m = 2$. Panels (f) and (g): the power and instability growth rate vs. the propagation constant for the vortex solitons in the first ring of potential (2) with $k = 6$. Panels (h) and (i): the simulated propagation of vortex solitons with m = 1 (h) and m = 2 (i) with $\mu = -10.34$ and $k = 6$.

Comparing Figs. 3(e), 3(f) and 4(f), 4(g), it is found that the corresponding solitons residing in the first ring with $m = 0$ [Fig. 3(f)] and $m = 1$ [Fig. 4(g)] manifest instability in the entire existence regions, whereas the solitons with $m = 2$ [Fig. 4(g)] have a stability region at $\mu < -10.34$. The simulated propagation of the unstable and stable solitons shown in Figs. 4(b) and 4(d) is displayed, respectively, in Figs. 4(h) and 4(i), to corroborate their (in)stability. This means that the vortex solitons with higher topological charges may be *more stable*, in contrast to the known properties of OVBs in previously studied systems, cf. Refs. [27, 65]. This phenomenon arises because the increase in the topological charge increases OAM carried by the vortex solitons, consequently enlarging the phase difference between adjacent "beads" in the necklace structure. For the topological charge $m = 2$, the phase difference attains $2\pi/3$, exceeding the critical value of $\pi/2$, above which the interaction between adjacent "beads" is repulsive, which helps to construct stable necklace solitons. On the other hand, for the solitons trapped in the second and third rings, the large spatial separation between the "beads" renders the interaction among them significantly weaker, tending towards their independent behavior. Therefore, the topological charge of these solitons hardly affects their existence and stability characteristics.

Further, vortex solitons with the azimuthal-modulation index $k = 8$ are presented in Fig. 5. It is observed that, for different topological charges $m$, the existence region of the solitons splits in two disjoint parts [Fig. 5(a)]. In the first part, the lattice can capture the OVB with components located in both the first and second rings, forming two-ring solitons, which carry zero topological charge $m = 0$ [see Fig. 5(c)] or $m = 1$ and $m = 2$ (not shown here). Owing to its deeper potential trough, the first ring captures a higher soliton power than the second one. When the total power of the solitons exceeds a certain critical value [Fig. 5(a)], the system creates new soliton states with $m = 0$, $m = 1$, and $m = 2$, whose power is concentrated in the second ring, as seen in Figs. 5(d)-5(f).

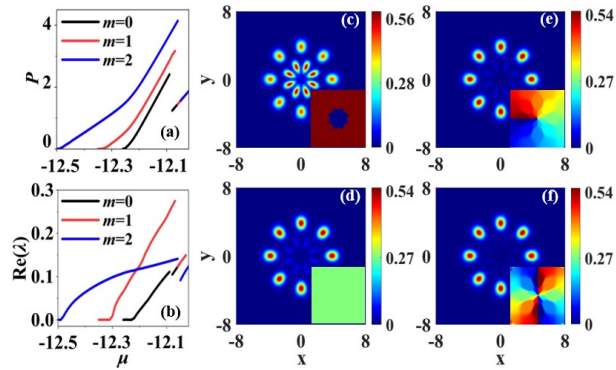

Fig. 5. The power (a) and instability growth rate (b) vs. the propagation constant for octupole vortex solitons with different topological charges in the first and second rings with $k = 8$ in potential (2). The power-density and phase distributions of solitons are displayed for $\mu = -12.15$, $m = 0$ (c), $\mu = -12.05$, $m = 0$ (d), $\mu = -12.05$, $m = 1$ (e), and $\mu = -12.05$, $m = 2$ (f).

Compared to Fig. 3(b), due to a smaller azimuthal-modulation period in Fig. 5(d), the solitons are trapped in the second ring, with a small portion penetrating into the first ring. Specifically, when $m = 2$, the phase difference between the adjacent "beads" attains $\pi/2$, leading to the zero net force between them. However, for $m = 1$, the phase difference is smaller than $\pi/2$, the net inter-bead interaction is attractive, which induces a trend for inward contraction of the necklace structure. Therefore, in comparison to the case of $m = 2$, additional components are more prone to be generated in the first ring when $m = 1$, see Figs. 5(e) and 5(f).

The propagation dynamics of octupole necklace OVBs with topological charges $m = 0$ and 2 are presented in Figs. 6(a) and 6(b). Driven by the competition between the diffraction and self-focusing, oscillations of the solitons are observed between the first and second rings in Figs. 6(a)-6(d)]. Unlike the conclusion drawn for $k = 6$, here the increase of the topological charge renders the solitons unstable. This happens because, at $k = 6$, the phase differences between "beads" within the single-ring solitons satisfy the stability condition, whereas, at $k = 8$, the oscillatory behavior between the "beads" in the second ring and the first ring disrupts the phase differences among "beads" within the same ring. When the topological charge is nonzero, the interaction between individual "beads" becomes complex due to the phase difference between them, thereby rendering solitons unstable. As shown in Fig. 6(b), for $m = 2$, although the soliton initially exhibits a breathing-like behavior during the early stage of the propagation, its instability rapidly grows at the later stage, ultimately destroying the breathing dynamics.

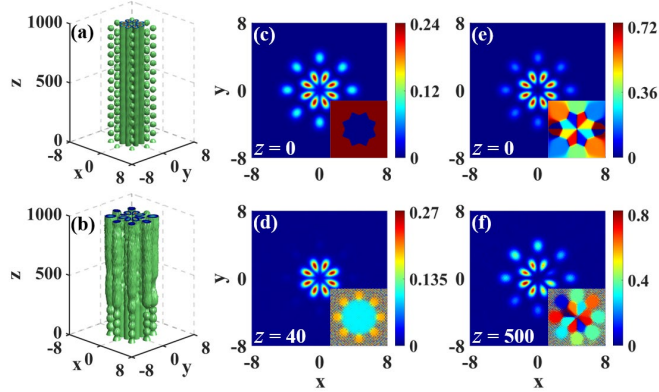

Fig. 6. The propagation dynamics of octupole necklace vortex solitons with $m = 0$ (a) and $m = 2$ (b) with $\mu = -12.23$ and $k = 8$ in the potential (2). Panels (c), (d) and (e), (f) are power-density distributions at different propagation distances, indicated in the panels, which correspond to (a) and (b), respectively. The insets in (c)-(f) represent the phase structure of the solitons.

Finally, the multipole necklace solitons with the azimuthal-modulation index $k = 12$ in potential (2) and a high topological charge, $m = 3$, are shown in Fig. 7. With the propagation constant $\mu \leq -13.66$ [Fig. 7(a)], the soliton is distributed in both the third and second rings [see Fig. 7(c)], and demonstrates stable propagation over long distances, see Figs. 7(d) and 7(g). Conversely, with the propagation constant $\mu > -13.66$, the soliton overflows from the second ring into the third one [Fig. 7(e)], and exhibits unstable breathing dynamics in Figs. 7(f) and 7(h). These conclusions, concerning the stability and instability of the solitons, are corroborated by the eigenvalue spectra, as presented in Figs. 7(b1) and 7(b2). This approach, utilizing the multi-ring fragmented arrangements to generate stable necklace solitons with high topological charges, suggests a novel pathway for enhancing the capacity of transmission channels.

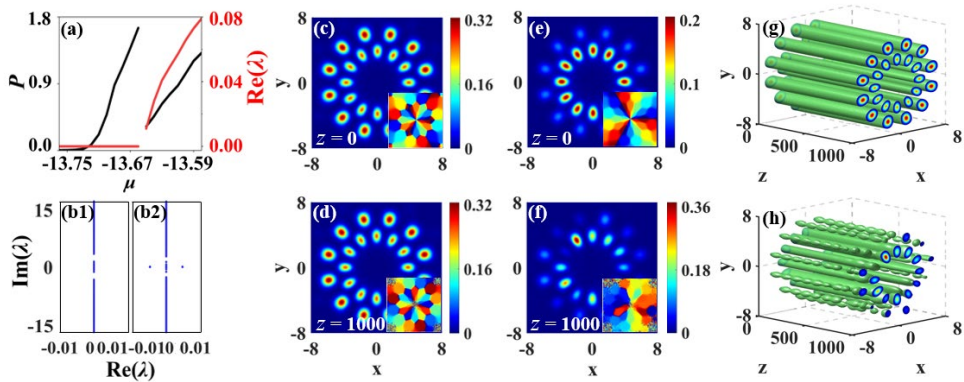

Fig. 7. (a) The power and instability growth rate vs. the propagation constant for multipole necklace solitons with the azimuthal-modulation index $k = 12$ in the potential (2) and $m = 3$. (b1, c, d, g) Stable and (b2, e, f, h) unstable multipole solitons with $k = 12$, $m = 3$, $\mu = -13.67$ and $k = 12$, $m = 3$, $\mu = -13.65$, respectively. (b1, b2) The respective eigenvalue spectra. (c, e) The power-density distributions at $z = 0$; (d, f) at $z = 1000$; (g, h) the stable and unstable propagation dynamics.

## 4. Conclusion

This work addresses the existence, stability and propagation dynamics of necklace vortex solitons in the self-focusing Kerr media with the azimuthally modulated (fragmented) Bessel lattice potentials, defined as per Eq. (2). The model is relevant in the spatial domain for the optical bulk waveguide, as well as in the spatiotemporal domain for optical cavities, the same model is also relevant, as the spatiotemporal one, for the effectively two-dimensional BEC. Due to the multi-ring structure of the Bessel lattice, solitons can be trapped in each ring (annular potential trough) or two adjacent ones. In the case of single-ring solitons, the existence region is broader, in terms of the power, for the inner ring. As for the multi-ring solitons, octupole necklace solitons are trapped in the first and second rings. The stable 12-pole necklace vortex solitons are also observed in the second and third rings. Notably, the propagation of the zero-vortex octupole necklace solitons exhibits stable breathing-like dynamics, with the power flow between adjacent "beads" of the necklace structure in the first and second rings. Unlike the conventional properties of OVBs, the increase of the topological charge *enhances* the stability of the multipole solitons confined to the first ring, in a certain parameter region. The stability of higher-*m* vortex solitons is closely related to their ring numbers and the azimuthal modulation parameter, for a given value of the potential depth. As the azimuthal modulation parameter increases, the location of the multi-ring solitons shifts toward rings with larger radii, thereby making the vortex solitons with higher topological charges stable in a certain range of the propagation constant. This study provides the theoretical basis for novel schemes of optical-field manipulations in nonlinear systems, and offers potential applications to optical communication and data processing.

**Funding.** This work was supported by the National Natural Science Foundation of China (Grant No. 62575165, 62305199), the Natural Science Foundation of Shanxi Province (Grant No. 202203021221016) and Shanxi graduate Education Innovation Program (Grant No. 2024YZ06).

**Disclosures.** The authors declare that there are no conflicts of interest related to this article.

**Data availability.** Data underlying the results may be obtained from the authors upon reasonable request.